\documentclass{aa}
\usepackage{epsfig}

\def\etal{{\rm et al.\thinspace}}
\def\eg{{\rm e.g.\ }}
\def\etc{{\rm etc.\ }}
\def\ie{{\rm i.e.\ }}
\def\cf{{\rm cf.\ }}

\def\spose#1{\hbox to 0pt{#1\hss}}
\def\ltsimm{\mathrel{\spose{\lower 3pt\hbox{$\sim$}}
	\raise 2.0pt\hbox{$<$}}}
\def\gtsimm{\mathrel{\spose{\lower 3pt\hbox{$\sim$}}
	\raise 2.0pt\hbox{$>$}}}
\def\Mdot{\hbox{${\dot M}$} \,}

\def\km{{\rm\thinspace km}}
\def\cm{{\rm\thinspace cm}}
\def\s{{\rm\thinspace s}}
\def\yr{{\rm\thinspace yr}}
\def\kmps{\hbox{${\rm\km\s^{-1}\,}$}}
\def\erg{{\rm\thinspace erg}}
\def\ergps{\hbox{${\rm\erg\s^{-1}\,}$}}
\def\Msol{\hbox{${\rm\thinspace M_{\odot}}$}}
\def\Msolpyr{\hbox{${\rm\Msol\yr^{-1}\,}$}}

\begin{document}
   
\title{In Hot Pursuit of the Hidden Companion of $\eta$~Carinae: An X-ray 
Determination of the Wind Parameters}

\author{J.M. Pittard\inst{1} \and M.F. Corcoran\inst{2,3}}

\institute{Department of Physics \& Astronomy, The University of Leeds, 
        Woodhouse Lane, Leeds, LS2 9JT, UK
 \and Universities Space Research Association, 7501 Forbes Blvd, Ste 206,
        Seabrook, MD 20706, USA
 \and Laboratory for High Energy Astrophysics, Goddard Space Flight Center,
        Greenbelt, MD 20771, USA}

\offprints{J. M. Pittard, \email{jmp@ast.leeds.ac.uk}}

\date{Received <date> / Accepted <date>}

\abstract{
We present X-ray spectral fits to a recently obtained 
{\em Chandra} grating spectrum of $\eta$~Carinae, one of the most massive and 
powerful stars in the Galaxy and which is strongly suspected to be a 
colliding wind binary system. Hydrodynamic models of colliding winds
are used to generate synthetic X-ray spectra for a range of 
mass-loss rates and wind velocities. They are then fitted against 
newly acquired {\em Chandra} grating data. We find that due to the low 
velocity of the primary wind 
($\approx 500 \kmps$), most of the observed X-ray emission appears to 
arise from the shocked wind of the companion star. We use the duration of
the lightcurve minimum to fix the wind momentum ratio at
$\eta = 0.2$. We are then able to obtain a good fit to the data by 
varying the mass-loss rate of the companion and the terminal velocity of 
its wind. We find that $\Mdot_{2} \approx 10^{-5} \;\Msolpyr$ and 
$v_{\infty_{2}} \approx 3000 \; \kmps$. With 
observationally determined values of $\approx 500-700 \; \kmps$ for the
velocity of the primary wind, our fit implies a primary mass-loss rate of 
$\Mdot_{1} \approx 2.5 \times 10^{-4} \;\Msolpyr$. This value is 
smaller than commonly inferred, although we note that a lower mass-loss 
rate can reduce some of the problems noted by Hillier \etal (\cite{HDIG2001}) 
when a value as high as $10^{-3} \;\Msolpyr$ is used. The wind parameters
of the companion are indicative of a massive star which may or may not be
evolved. The line strengths appear to show slightly sub-solar abundances,
although this needs further confirmation. Based on the over-estimation of 
the X-ray line strengths in our model, and re-interpretation of the HST/FOS 
results, it appears that the homunculus nebula was produced by the 
primary star.
}

\titlerunning{In Hot Pursuit of the Hidden Companion of $\eta$~Carinae}
\authorrunning{Pittard \& Corcoran}

\maketitle

\label{firstpage}

\section{Introduction}
\label{sec:intro}
The superluminous star $\eta$~Carinae (HD~93308, HR~4210) continues to 
be extensively studied over a host of different wavelengths, yet remains 
intriguingly enigmatic. It is amongst the most unstable stars known. 
In the 1840s, and again in the 1890s, it underwent a series of giant 
outbursts (\eg Viotti \cite{V1995}) which ejected large 
masses of material into the surrounding 
medium. {\em HST} images of the resulting bipolar nebula (\eg Morse \etal
\cite{MDB1998}), known as the Homunculus, show it to be amongst the most 
spectacular in our Galaxy. The central object is now largely obscured by
dust, and the cause of the outbursts and the nature
of the underlying star (at outburst and today) remain speculative. 
The source continues to show brightness fluctuations and emission-line 
variations. Further details of $\eta$~Car can be found in the review by 
Davidson \& Humphreys (\cite{DH1997}).
  
In recent years evidence for binarity in this system has been 
accumulating. Damineli (\cite{D1996}) first noted a 5.5~yr period in the 
variability of the He I 10830~${\rm \AA}$ line. Further photometric and
radial velocity studies (Damineli \etal \cite{DCL1997}, \cite{D2000}), 
X-ray observations (Tsuboi \etal \cite{TKSP1997}; Corcoran \etal 
\cite{CFPSD2000} and references therein), and radio {\mbox data} (Duncan \etal 
\cite{D1995}, \cite{DWRL1999}) have supported the 5.5~yr period and the binary 
hypothesis. However, the ground 
based radial velocity curve was not confirmed by higher resolution 
spectra with STIS, indicating that at least the time of periastron 
passage is not well defined by the UV and optical spectra (Corcoran 
\etal \cite{CISP2001}). A comparison of the 
abundances from the central object(s) and
the composition of the Homunculus nebula has also determined that there
are at least two stars in this system (Lamers \etal \cite{LLPW1998}).

$\eta$~Car is often classified as a luminous blue variable (LBV). 
These are massive
stars believed to be in a rapid and unstable evolutionary phase in which
many solar masses of material are ejected into the interstellar medium over
a relatively short period of time ($\approx 10^{4}$~yr). LBVs are regarded
as a key phase in the evolution of massive stars, during which a transition 
into a Wolf-Rayet star occurs (\eg Langer \etal \cite{L1994}; Maeder \& 
Meynet \cite{MM2001}). Due to their rarity and complex nature however, 
we unfortunately still have no definitive theory for mass-loss during the 
LBV stage. The majority of proposed mechanisms to drive LBV 
instabilities, the onset of higher mass-loss rates and underlying eruptions,
are concerned with the importance of radiation pressure 
within the outer envelope of the LBV, and
for example utilize pulsational instabilities (\eg Guzik \etal \cite{GCD1999}),
dynamical instabilities (\eg Stothers \& Chin \cite{SC1993}), or presuppose
Eddington-like instabilities. The latter could arise from an enhancement in
opacity as the star moves to lower temperatures (\eg Lamers \cite{Lam1997}),
or from the influence of rotation (\eg Langer \cite{Lan1997}; Zethson \etal
\cite{Z1999}). Alternatively, the possibility that binarity plays a 
fundamental role in explaining observed LBV outburst properties has also 
been considered (Gallagher \cite{G1989}), though most LBVs are not known 
binaries. Clearly, determining the wind and stellar properties of LBV
stars is paramount (see, for example, the discussions in Leitherer 
\etal \cite{L1994} and Nota \etal \cite{N1996}).

An important question is the degree to which binarity influences the 
properties of LBVs (\ie do LBVs in binaries evolve differently than single 
LBVs?). So while the presence of a companion can be exploited to help
measure the mass of such stars, we must bear in mind that {\mbox binary} LBVs
and single LBVs may be quite distinct objects. Therefore,
in order to use $\eta$~Car to understand some of the 
defining LBV characteristics such as their extremely high mass-loss 
rates, we first need to determine beyond all doubt that $\eta$~Car is in 
fact a binary, and then to determine the influence of the companion on 
the system.

\begin{figure*}
\begin{center}
\psfig{figure=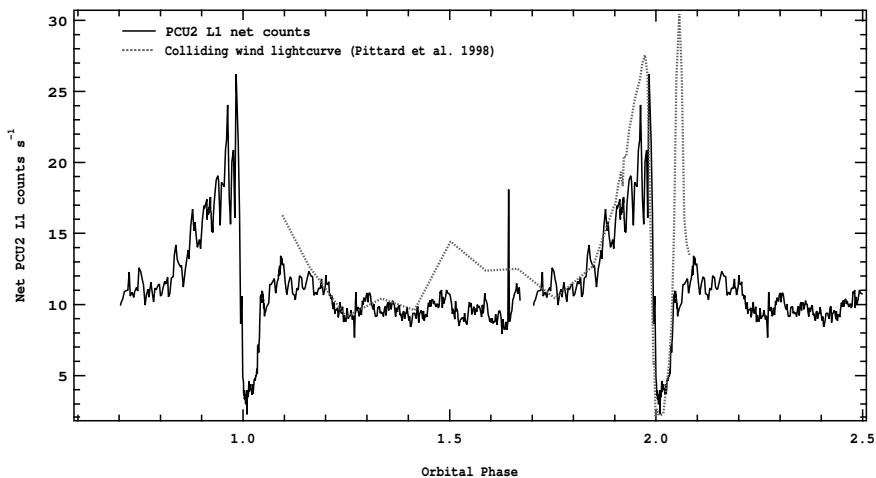,width=12.0cm}
\end{center}
\caption[]{Lightcurve of $\eta$~Car observed with the {\it RXTE} satellite 
and phased to the 5.5~yr orbital period. Plotted are counts detected in 
layer 1 of the second proportional counter unit (PCU2) and a predicted 
lightcurve (Pittard \etal \cite{PSCI1998}) from a numerical model of the 
wind-wind collision. The two agree well, particularly the duration of the 
minimum. The rise from minimum is not in good agreement, but this is thought 
to be due to the limitations of modelling the wind collision in 2D. The 
rapid change in position angle of the stars through periastron passage 
skews the shock cone which causes the line of sight in fully 3D models
to remain in the denser wind of the primary until later phases, increasing 
the absorption at these times (Pittard \cite{P2000})}.
\label{fig:eta_lc}
\end{figure*}

Investigations over the last few years have already helped to form a basic 
picture of $\eta$~Car. The orbital {\mbox parameters}, although uncertain,
indicate the presence of an early-type companion star, which will also have a
powerful stellar wind. In such binaries, a region of hot shocked gas 
with temperatures in excess of 10~million~K is created where the stellar 
winds collide (Prilutskii \& Usov \cite{PU1976}; Cherepashchuk \cite{C1976}). 
The wind-wind collision (WWC) region is expected to 
contribute to the observed emission from this system, particularly at 
X-ray and radio wavelengths. Previous X-ray observations revealed
extended soft emission from the nebula and strong, hard, highly absorbed, 
and variable emission closest to the star (Corcoran \etal \cite{C1995}; 
Weis \etal \cite{WDB2001}), 
in contrast to the emission characteristics from single stars, 
which are typically softer, much less absorbed, substantially weaker and 
relatively constant in intensity. Since 1996 Feb $\eta$~Car has been
continuously monitored by {\it RXTE} in the 2--10~keV band (\eg Corcoran 
\etal \cite{CISP2001}). The lightcurve (Fig.~\ref{fig:eta_lc}) 
{\mbox contains} remarkable detail showing a slow, almost linear, rise 
to maximum over a period of $\approx 1$~yr, followed by a rapid drop to  
approximately 1/6 of the peak intensity for $\approx 3$~{\mbox months}, 
an almost as sharp rise to approximately 1/2 of the peak 
intensity level, and then almost constant intensity for $\approx 3/4$ of 
the proposed 5.5~yr orbital period. The drop to minimum was successfully 
predicted from {\mbox numerical} models of the WWC 
(Pittard \etal \cite{PSCI1998}) before being actually observed. 

Small scale quasi-periodic outbursts in the X-ray lightcurve have also been
detected (Corcoran \etal \cite{CISDPS1997}). Estimates of changes in the
timescale between successive flares as a function of phase were made 
by Davidson \etal (\cite{DIC1998}) for a variety of assumed orbital elements.
{\em RXTE} X-ray observations obtained after the X-ray minimum
seem to show a lengthening of the flare timescale (Ishibashi \etal 
\cite{I1999}), which indirectly support the {\mbox binary} model. 
The latest published X-ray observation of $\eta$~Car is of a 
high resolution grating spectrum taken with
the {\it Chandra} X-ray observatory (Corcoran \etal \cite{C2001}). Preliminary
analysis has revealed the presence of strong forbidden line emission, which
suggests that the density of the hot gas, $n_{\rm e} < 10^{14} \cm^{-3}$. This
can be contrasted with the newly published X-ray grating spectra of the
single stars $\theta^{1}$~Ori~C (Schulz \etal \cite{SCHL2000}), 
$\zeta$~Ori (Waldron 
\& Cassinelli \cite{WC2001}), and $\zeta$~Pup (Kahn \etal \cite{K2001}), 
all of which have weak forbidden lines (indicative of either high densities
or high UV flux near the line forming region). This lends 
further support to a wind-wind collision model, although it is possible
that some of the forbidden emission may be related to the surrounding 
nebula.

If $\eta$~Car is in fact a 
binary system, the orbital {\mbox elements} and stellar parameters are not 
yet tightly constrained by current observations. Ground-based 
observations (for which good phase coverage exist) are hampered by 
poor spatial resolution and thus suffer contamination from strong 
nebular emission.  High-spatial resolution spectra have been 
obtained by STIS but phase coverage is currently very limited. {\em HST} STIS 
observations at two different phases of the 5.52 year cycle (Davidson 
\etal \cite{DI2000}) did {\em not} confirm the predicted variations in the 
radial {\mbox velocity} of the emission lines based on the ground-based radial 
{\mbox velocity} curve (Damineli \etal \cite{D2000}). If $\eta$~Car is a 
binary, it is vitally important to determine the stellar {\mbox parameters} 
of the companion so that the effect of the companion on observations can be 
understood and the correct stellar parameters of the primary can be 
derived.

\subsection{X-ray Emission from Colliding Winds}

The wealth of information contained in X-ray spectra of colliding wind
binaries (\eg the density, temperature, {\mbox velocity}, abundance, and 
distribution of the shocked gas in the wind collision region) 
has been a strong motivating force for observers and theorists alike 
in this field.
Since the hot plasma in most colliding wind shocks is optically thin and 
collisionally ionized, and is generally assumed to be in collisional 
equilibrium, Raymond-Smith (Raymond \& Smith \cite{RS1977}) or
MEKAL (Mewe \etal \cite{MKL1995}) spectral models  
are normally fitted to such data (\eg Zhekov \& Skinner \cite{ZS2000};
Rauw \etal \cite{RSPC2000}; Corcoran \etal \cite{C2001}). However, 
the multi-temperature, multi-density nature of the WWC region
means that at best simple fits with one- or two-temperature Raymond-Smith 
models can {\mbox only} {\em characterize} the broad properties of the 
emission. 
In this way one can estimate an `average' temperature of the shocked gas, 
and an `average' absorbing column, but little is learned of the underlying 
stellar wind parameters. At worst the application of one- or two-temperature
models to what is inherently multi-temperature emission can lead to 
spurious values of some of the fit parameters, \eg abundances (\cf
Strickland \& Stevens \cite{SS1998}).

Complex numerical hydrodynamical models have often been applied 
to gain insight into colliding wind {\mbox systems} (\eg Stevens \etal 
\cite{SBP1992}, Pittard \etal \cite{PSCI1998}). However, while undoubtedly 
useful, their interpretation can be {\mbox difficult}, and to date 
there have been only  
two published papers where observed spectra are {\em directly fitted} with 
synthetic spectra from such models. In the pioneering work of Stevens \etal 
(\cite{S1996}), medium resolution {\it ASCA} spectra of the Wolf-Rayet binary 
$\gamma^{2}$ Velorum were fitted against a {\em grid} of synthetic spectra.
In this fashion they were able to obtain direct estimates of the 
mass-loss rates and terminal velocities of the individual stellar winds. 
As mass-loss rates obtained from measures of radio flux or spectral line
fits depend on a variety of untested assumptions, the importance of a new 
independent method to complement estimates from free-free radio or sub-mm 
observations, or from H$\alpha$ or UV spectral line fitting, cannot be 
stressed enough. Rates derived by Stevens \etal (\cite{S1996}) 
with this new method were significantly lower than the commonly accepted
estimates for $\gamma^{2}$ Velorum based on radio observations, but an
indication of the future benefits of this method was realized when
both sets of estimates were {\mbox later} brought into agreement 
following a surprisingly large reduction in the distance to this 
star from {\it Hipparcos} {\mbox data}\footnote{The thermal radio flux, 
$S_{\nu} \propto \Mdot^{4/3} D^{-2}$ (where $D$ is the distance to the 
source) whereas the X-ray flux from
an adiabatic wind collision is $F_{\nu} \propto \Mdot^{2} D^{-2}$. Therefore
$\Mdot_{\rm radio} \propto D^{3/2}$ whereas $\Mdot_{\rm xray} \propto D$. 
If $D$ is revised downwards, $\Mdot_{\rm radio}$ decreases faster than
$\Mdot_{\rm xray}$.}. We note that this method can also provide insights 
into the values of parameters which are otherwise virtually impossible 
to estimate, such as the mass-loss rate of the companion, $\Mdot_{2}$,
or the characteristic ratio of the pre-shock electron and ion temperature
(Zhekov \& Skinner \cite{ZS2000}).
The quality of recently available X-ray grating spectra now gives us 
access to important X-ray emission line diagnostics which should 
severely constrain models of the X-ray emission distribution. This means
that stellar wind parameters can in principle be reliably estimated from 
analysis of X-ray grating spectra of colliding wind binaries. 

In addition to testing the binary hypothesis, 
the {\it Chandra} grating spectrum of $\eta$~Car (Corcoran \etal 
\cite{C2001}) provides the ideal opportunity to test the method 
developed by Stevens \etal (\cite{S1996}) against a spectrum of much higher 
spectral resolution, and to pin down important physical parameters of 
the system. In this paper we therefore fit the X-ray grating spectrum 
using a grid of colliding wind emission models to i) test the binary
hypothesis, and ii) to attempt to obtain accurate estimates of the wind 
{\mbox parameters} of each star. The fact that we are able to obtain good fits,
with sensible model parameters, gives us further confidence in the binary
hypothesis. We also find that unlike the
UV and optical work where fits to the primary are made difficult by 
significant contamination from the companion star, instead 
the X-ray emission arises from the shocked wind of the {\em companion} and 
suffers essentially zero contamination from the wind of the primary. 
Therefore the X-ray data uniquely samples {\mbox parameters} of the companion,
in contrast to the optical analysis which probes the nature of the primary.
In this sense our analysis is entirely complementary to the complex 
fits to the UV and optical HST spectrum of $\eta$~Car by Hillier \etal 
(\cite{HDIG2001}). Our analysis also provides us with a new estimate 
of the mass-loss rate of the primary {\mbox star}. For details of the 
{\it Chandra} observation and an initial analysis of the data the reader is
referred to Corcoran \etal (\cite{C2001}). Here we only note that 
there is no significant contamination of the dispersed spectrum from 
any spatially resolved emission (\ie in the Homunculus). In 
Section~\ref{sec:syn_spec}
we discuss the creation and variation of the model spectral grid; in 
Section~\ref{sec:fits} we describe the fit results; and in 
Section~\ref{sec:conclusions} we summarize and conclude.

\section{The Synthetic Colliding Wind Spectra}
\label{sec:syn_spec}
The method applied by Stevens \etal (\cite{S1996}) as discussed in 
Section~\ref{sec:intro} is as follows. We first calculated a whole host of 
hydrodynamical models with different stellar wind {\mbox parameters}. We then 
generated a synthetic X-ray spectrum from each model. Finally, we 
fit the grid of synthetic spectra to the actual data. 
Best-fit values are then obtained 
for each parameter of interest (\eg $\Mdot_{1}$, $\Mdot_{2}$, 
$v_{\infty_{1}}$, $v_{\infty_{2}}$ \etc). 
Specific points in relation to this method are highlighted below. 

\subsection{Hydrodynamical Models of $\eta$~Carinae}
\label{sec:hydro_models}
The colliding wind models were calculated using {\sc VH-1} (Blondin \etal 
\cite{BKFT1990}), a Lagrangian-remap version of the third-order accurate
Piecewise Parabolic Method (PPM; Colella \& Woodward \cite{CW1984}).
The stellar winds are modelled as ideal gases with adiabatic index 
$\gamma = 5/3$. Radiative cooling is included via the method of operator
splitting, and is calculated as an optically thin plasma in ionization
equilibrium. The cooling curve for the temperature range $4.0 < 
{\rm log}\; T < 9.0$ was generated using the Raymond-Smith plasma code.
Despite some evidence to the contrary (see Corcoran \etal \cite{C2001}), 
we have used solar abundances throughout this work, since it keeps the 
problem as simple as possible in this first investigation. We aim to 
model non-solar abundances in a later paper.

The simulations were calculated assuming cylindrical symmetry of the 
wind collision zone - orbital effects are negligible at the phase of 
the {\it Chandra} observation ($\phi=0.60$). Grid sizes spanned a range 
from $300\;{\rm x}\;300$ to $550\;{\rm x}\;450$ 
cells. Each grid cell was square and of constant size on an individual 
grid. The linear dimension of each cell was either $2.67 \times 10^{12}$ or 
$4.0 \times 10^{12}$~cm. The maximum  distance from the axis of 
symmetry was $1.2 \times 10^{15}$~cm in all cases. At the
phase of the observation, the orbital separation of the 
stars using the ephemeris of Corcoran \etal (\cite{CISP2001}) is  
$\approx 4 \times 10^{14}$~cm. Thus there is either 100 or 150 grid cells 
between the stars. The large separation means that the 
winds are likely to collide at very near their terminal
velocities. We therefore assume that we can treat the winds as being
instantaneously accelerated to their terminal velocities, and do not 
consider any radiative driving effects. In comparison with the work on 
$\gamma^{2}$~Velorum (Stevens \etal \cite{S1996}), assuming 
terminal velocity winds and negligible binary rotation 
is more valid for $\eta$~Car. We further assume that the winds are 
spherically symmetric.

To compute a grid of synthetic spectra we initially varied
four parameters ($\Mdot_{1}$, $\Mdot_{2}$, 
$v_{\infty_{1}}$, $v_{\infty_{2}}$) in the {\mbox hydrodynamic} models. 
During our investigation we also included the separation of the stars, $D$,
as an additional free parameter. From fits at this point it was found 
that $D \approx 4 \times 10^{14}$~cm with a small uncertainty, in good 
agreement with the expected value from our current understanding of the 
orbit. We therefore fixed it at this value for the rest of our analysis. 
However, we found that there were large uncertainties on the values 
of $\Mdot_{1}$ and $v_{\infty_{1}}$. With the benefit of hind-sight this is not
too surprising given the known slow speed of the primary wind 
($v_{\infty_{1}} \approx 500-700 \;\kmps$), which means that the shocked 
primary wind is not a strong source of X-ray emission at hard energies.
Therefore for our final grid we fixed the terminal velocity of the 
primary star at $500\;\kmps$ and adjusted $\Mdot_{1}$ to obtain a 
desired value for the wind momentum ratio, 
$\eta = \Mdot_{2}\;v_{\infty_2}/(\Mdot_{1}\;v_{\infty_1})$, 
with $\Mdot_{2}$ and $v_{\infty_2}$ as free parameters. 

The range of the free parameters used in our final grid is given in 
Table~\ref{tab:hydro_grid}. To restrict the number of 
models to a manageable number the parameter steps are fairly coarse.
In future papers we will use a much finer grid.
The range in the value of $\eta$ corresponds to either the
wind of the primary ($\eta=0.1$) or the secondary ($\eta=5$)
dominating, or the winds having equal momentum fluxes ($\eta=1$).
The distance of the stagnation point from the centre of the primary 
star is given by

\begin{equation}
\label{eq:rob}
r = \frac{1}{1 + \sqrt{\eta}} \; D,
\end{equation}

\noindent and ranges from $1.2 \times 10^{14}$~cm to $3.0 \times 10^{14}$~cm. 
In the most extreme cases, this is still of order ten to a hundred times the 
radius of the stars, and justifies our use of terminal velocity winds.
The kinetic power of the primary wind ranged from 
$4.7 \times 10^{34} \;\ergps$ to $7.9 \times 10^{38} \;\ergps$, 
and from $7.1 \times 10^{35} \;\ergps$ to $7.9 \times 10^{38} \;\ergps$
for the secondary wind. The combined kinetic power of the winds ranges
from $7.6 \times 10^{35} \;\ergps$ to $1.6 \times 10^{39} \;\ergps$.
The half-opening angle of the contact discontinuity
measured from the line between the secondary star and
the shock apex ranges from $\approx 50-120^{\circ}$ (\cf Eichler \& Usov 
\cite{EU1993}). 

The effect of radiative cooling can be quantified by the
parameter $\chi$, the ratio of the cooling timescale to the dynamical 
timescale of the system. For shocked gas near the local minimum in the 
cooling curve at $T \approx 2 \times 10^{7}$~K,

\begin{equation}
\label{eq:chi_sbp}
\chi \approx \frac{v_{8}^{4} d_{12}}{\Mdot_{-7}},
\end{equation}

\noindent where $v_{8}$ is the wind velocity in units of $1000 \;\kmps$,
$d_{12}$ is the distance to the contact discontinuity in {\mbox units} 
of $10^{12}$~cm,
and $\Mdot_{-7}$ is the mass-loss rate in units of $10^{-7} \; \Msolpyr$
(\cf Stevens \etal \cite{SBP1992}). This equation is valid for post-shock 
temperatures in the range $10^{7} \ltsimm T \ltsimm 10^{8.2}$~K 
($680/\sqrt{\mu} \ltsimm v \ltsimm 2700/\sqrt{\mu} \; \kmps$ where 
$\mu {\rm m_{H}}$ is the average particle mass in grammes). For material
with solar abundance ($\mu = 0.6$), this corresponds to 
$0.9 \ltsimm v_{8} \ltsimm 3.5$. For winds with slower velocities, 
the shocked gas lies on the negative slope of the cooling curve. In this 
case the dependence of $\chi$ on the velocity of the wind is steeper:

\begin{equation}
\label{eq:chi_2}
\chi \approx \frac{v_{8}^{5.2} d_{12}}{\Mdot_{-7}}.
\end{equation}

\begin{figure*}
\begin{center}
\psfig{figure=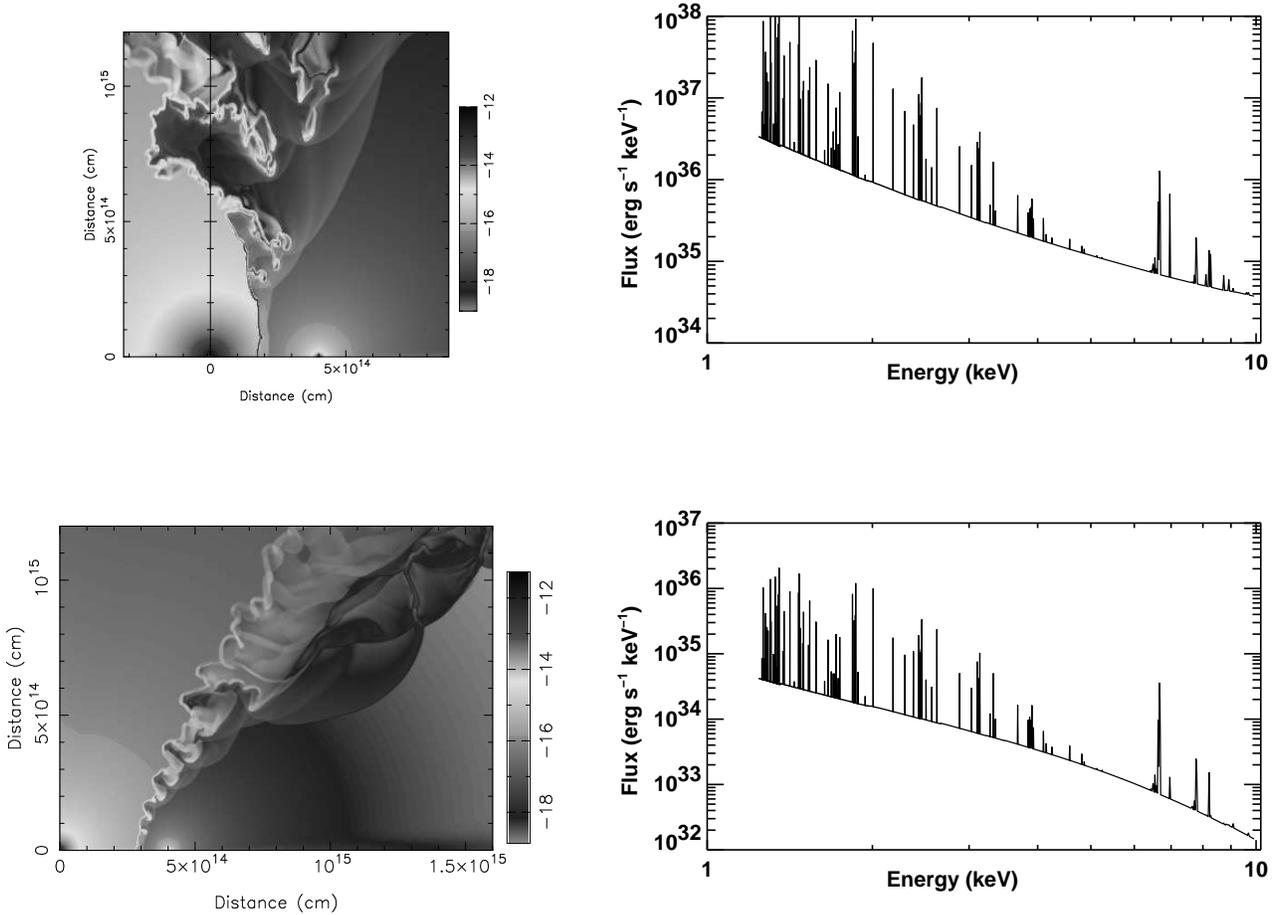,width=16.8cm}
\end{center}
\caption[]{Hydrodynamic simulations and theoretical spectra of the 
colliding winds in $\eta$~Car. The top panels show the results from
a model where the momentum of each wind is equal ($\eta=1$), whereas in 
the bottom panels the primary wind has a momentum 10 times greater 
than the secondary wind ($\eta=0.1$). In both cases a density plot 
(${\rm g\;cm^{-3}}$) from the 
hydrodynamical grid is shown on the left and the resulting intrinsic 
X-ray spectrum shown on the right (no absorption).
On each hydrodynamical grid the primary star is located at (0,0) and 
the secondary star is at ($4.0 \times 10^{14}$,0).
The shape and structure of the shocked region varies with the momentum
balance between the winds and the respective value of $\chi$ for each 
shocked wind. The full parameters used in these models were: upper panels - 
$\Mdot_{1} = 10^{-3} \;\Msolpyr$, $\Mdot_{2} = 10^{-4} \;\Msolpyr$, 
$v_{\infty_{1}} = 500 \; \kmps$, $v_{\infty_{2}} = 5000 \; \kmps$; 
lower panels - $\Mdot_{1} = 3 \times 10^{-4} \;\Msolpyr$, 
$\Mdot_{2} = 10^{-5} \;\Msolpyr$, $v_{\infty_{1}} =  500 \;\kmps$, 
$v_{\infty_{2}} = 1500 \;\kmps$. The luminosity and shape of the X-ray spectrum
are a direct consequence of these parameters.}
\label{fig:hydro}
\end{figure*}

\noindent In our grid of simulations, the shocked wind of the {\mbox primary}
star is 
almost always strongly radiative ($\chi_{1,\rm min} << 1.0$) while the 
secondary's is never strongly radiative ($\chi_{2,\rm min} = 2.0$). However, 
for some choices of the 3 parameters in Table~\ref{tab:hydro_grid}, the 
primary's wind can approach the point where it starts to become adiabatic 
($\chi_{1,\rm max} \approx 1.0$). In Fig.~\ref{fig:hydro}
we show two examples of the hydrodynamic calculations used to calculate 
synthetic spectra. Due to the difference in terminal velocity of the
two winds, the velocity shear at the wind interface always generates
Kelvin-Helmholtz instabilities in our models. The rapid cooling of at least
one of the two winds means that thin-shell instabilities also always
occur in our models. The colliding winds region is most unstable when
both shocked winds rapidly cool (\cf Stevens \etal \cite{SBP1992}).

\begin{table}
\begin{center}
\caption{Our final grid of stellar wind parameters for the 
hydrodynamical models used to generate synthetic spectra. The models
fix the terminal velocity of the primary wind at $500\;\kmps$ and the
stellar separation at $D=4.0 \times 10^{14}$~cm. The mass-loss rate
of the primary star is given by $\Mdot_{1} = \Mdot_{2}\;v_{\infty_2}/
(500 \;\eta)$.}
\label{tab:hydro_grid}
\begin{tabular}{llll}
\hline
Parameter & 1 & 2 & 3 \\
\hline
$\eta$ & 0.1 & 1.0 & 5.0 \\
$\Mdot_{2} \; (\Msolpyr)$ & $10^{-6}$ & $10^{-5}$ & $10^{-4}$ \\
$v_{\infty_{2}} \; (\kmps)$ & 1500 & 3000 & 5000 \\
\hline
\end{tabular}
\end{center}
\end{table}

The spectrum from each hydrodynamic model was {\mbox averaged} over 3 
`snapshots' each spaced by about 70~d. This is of order the wind dynamical
timescale and serves to approximate a time-averaged spectrum. It is
particularly important to adopt this approach for models where the 
wind collision region is very unstable since in these cases the flux can 
vary by greater than $\pm 50\%$ from one snapshot to the next. As already
noted by Corcoran \etal (\cite{C2001}) no variability is seen during the
$\approx 90\; {\rm ksec}$ exposure of the {\it Chandra} grating 
spectrum. This is to be expected, however, since the dynamical timescale 
is much longer. Two examples of `time-averaged' synthetic spectra are shown in 
Fig.~\ref{fig:hydro}, next to density plots of the corresponding
{\mbox hydrodynamic} calculation.

The synthetic spectra were calculated over the {\mbox energy} range 
1.26--10~keV
(10--1.26~$\AA$ respectively), and have a resolution of 0.005~$\AA$. The 
actual grating spectrum shows significant absorption below 1.5~keV and 
contains essentially no useful information below our lower limit of 1.26~keV. 
The spectral resolution of our synthetic spectra is approximately twice as 
high as the grating data that we model. Use of the Raymond-Smith plasma code 
implicitly assumes that the plasma is in thermal equilibrium. This is
generally true for colliding wind binary systems (see Luo \etal 
\cite{LMM1990}) and is a good assumption for $\eta$~Car.

\begin{figure*}
\begin{center}
\psfig{figure=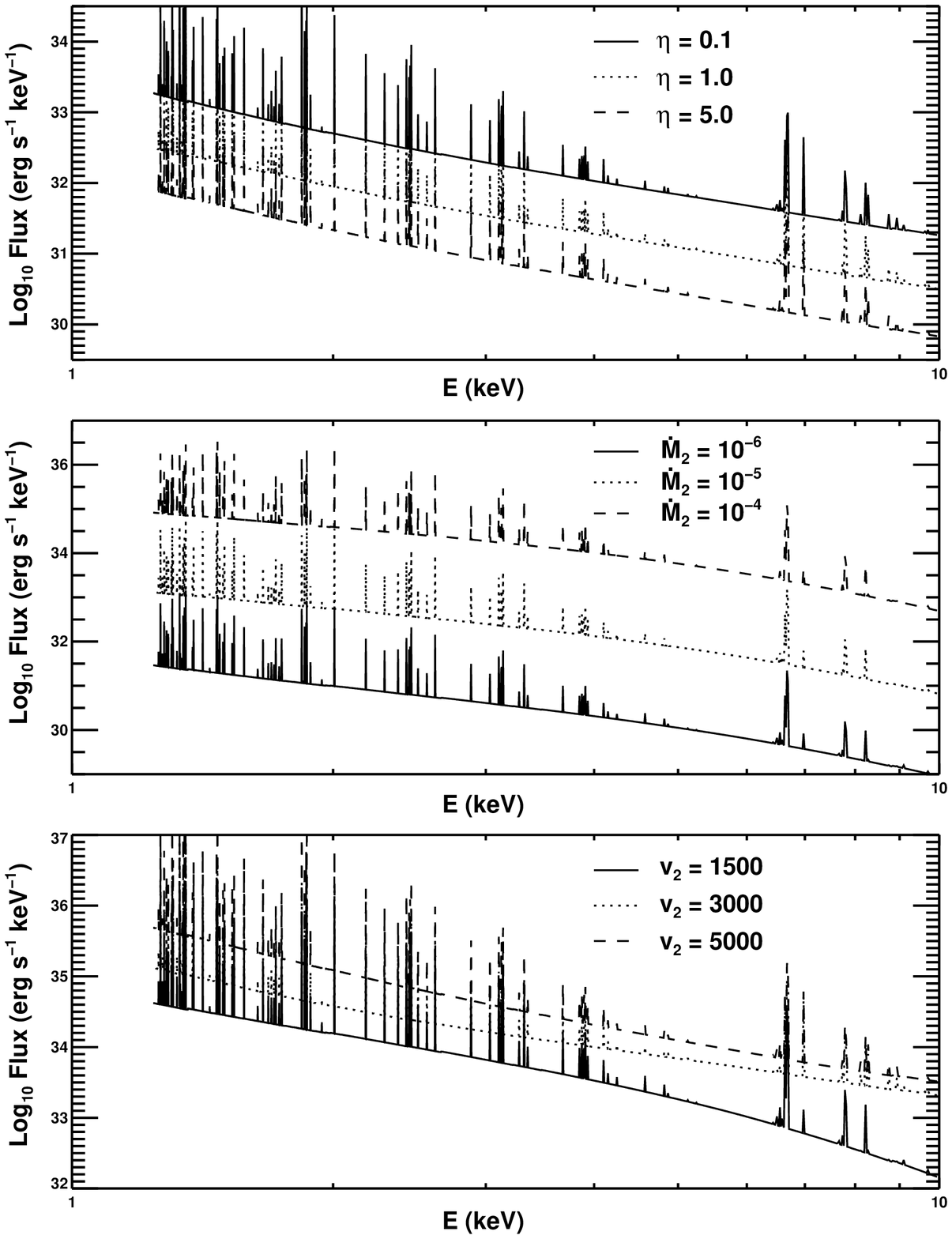,width=13.5cm}
\end{center}
\caption[]{Variation in the spectral shape and normalization of the 
theoretical colliding winds spectra in $\eta$~Carinae. 
The top panel shows the variation with $\eta$
($\Mdot_{2} = 10^{-6}\;\Msolpyr$, $v_{\infty_{2}} = 3000\;\kmps$).
The middle panel shows the variation with $\Mdot_{2}$ ($\eta = 5.0$, 
$v_{\infty_{2}} = 1500\;\kmps$).
The bottom panel shows the variation with $v_{\infty_{2}}$
($\eta = 0.1$, $\Mdot_{2} = 10^{-5}\;\Msolpyr$). 
}
\label{fig:varspec}
\end{figure*}

\subsection{Variations with $\eta$, $\Mdot_{2}$, and $v_{\infty_2}$}
\label{sec:syn_spec_var}
In Fig.~\ref{fig:varspec} and the following subsections we reveal
how the shape and normalization of the calculated spectra depend on 
the various free parameters of the grid.  

\subsubsection{Variation with $\eta$}
\label{sec:var_eta}
Due to the low preshock velocity of the primary wind, the X-ray emission from
the WWC is almost entirely from the shocked secondary wind. A greater fraction
of the secondary wind passes into the wind collision zone as the
relative momentum of the primary wind is increased, so we expect the
luminosity to increase as $\eta$ decreases, as indeed the top panel 
of Fig.~\ref{fig:varspec} shows. An important point,
however, is that the shape of the spectra do not seem dependent on the
value of $\eta$, contrary to our expectations before starting this
study.

\subsubsection{Variation with $\Mdot_{2}$}
\label{sec:var_mdot2}
As we increase the mass-loss rate of the secondary star in our models, 
the kinetic power and the density of its wind also increase. 
Therefore, for a given wind momentum ratio, the rate of 
conversion of kinetic energy to thermal energy increases, as does the
density of the postshock gas. Thus there is more available energy to radiate 
and a greater {\mbox efficiency} of radiation. Hence we should expect the X-ray
luminosity to scale strongly with increasing $\Mdot_{2}$. This is indeed 
what we see in the middle panel of Fig.~\ref{fig:varspec}, where an
order of magnitude increase in $\Mdot_{2}$ leads to an almost two order
of magnitude increase in X-ray luminosity, indicating that the 
calculations in this panel are in the adiabatic regime 
($L_{x} \propto \Mdot^{2}$; see 
Stevens \etal \cite{SBP1992}). However, there is no evidence for a 
softening of the spectrum with increasing $\Mdot_{2}$. This is
again contrary to our initial expectations (note, however, the discussion
in the next section).

\subsubsection{Variation with $v_{\infty_2}$}
\label{sec:var_v2}
If the preshock velocity of a wind is increased, the postshock temperature
increases ($T \propto v^{2}$). Therefore we expect the model spectra to harden
as the preshock velocity of the secondary wind is increased. Because the
kinetic power of the wind increases we also expect the {\mbox luminosity}
to increase. The variation of the spectrum with $v_{\infty_2}$ is shown 
in the bottom panel of Fig.~\ref{fig:varspec}. Here we find that our
expectations are only partially met. The overall {\mbox luminosity} does 
increase with $v_{\infty_2}$, and the spectrum does harden between 
$v_{\infty_2} = 1500-3000\;\kmps$, but it softens between
$v_{\infty_2} = 3000-5000\;\kmps$. An investigation into this surprising 
finding offers the following explanation.

For fixed $\eta$ and $\Mdot_{2}$, and variable $v_{\infty_2}$, the 
density of the wind varies as $\rho \propto \Mdot/v_{\infty_2}$. If
$v_{\infty_2}$ is increased $\rho$ decreases, which reduces the radiative
efficiency of the hot shocked gas near the stagnation point 
($\dot{E} \propto n^{2}$), which acts to slightly suppress the hard flux.
A plot of the average density against $T$ for 
$v_{\infty_2} = 1500,\;3000,\;5000 \;\kmps$ also reveals that for 
$10^{6.3} < T < 10^{7.4}$ the average density is higher when 
$v_{\infty_2}=5000\;\kmps$ than when $v_{\infty_2}=3000\;\kmps$, which tends
to keep up the soft emission relative to the hard emission. This appears
to be because there is some mixing at the interface
between the primary and secondary wind, and the shocked primary wind is
denser when $v_{\infty_2}=5000\;\kmps$ ($\Mdot_{1} = 3\times10^{-4},\;
{\mbox 6\times10^{-4}},\;10^{-3}\;\Msolpyr$ for 
$v_{\infty_2}=1500,\;3000,\;5000\;\kmps$). 
The mixed gas tends to be at these intermediate temperatures.

\begin{figure}
\begin{center}
\psfig{figure=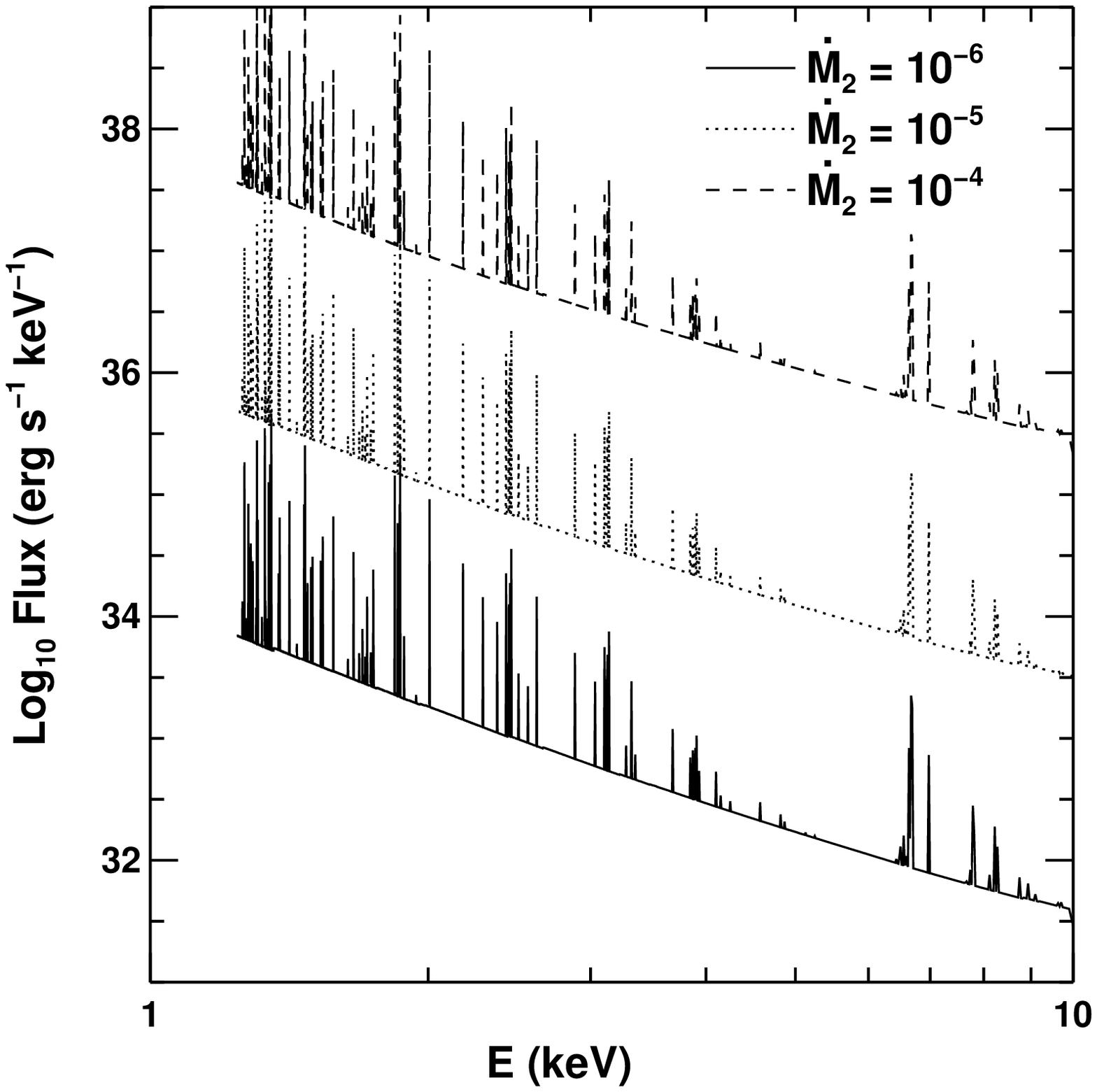,width=7.8cm}
\end{center}
\caption[]{Variation in the spectral shape and normalization for a very fast
secondary wind ($v_{\infty_{2}} = 5000\;\kmps$) and a fixed wind momentum
ratio ($\eta=0.1$). 
As $\Mdot_{2}$ increases the spectrum becomes slightly harder.}
\label{fig:mdot2_var}
\end{figure}

The overall net effect is that the emission from the hot gas is
suppressed relative to the cooler gas, which softens the
spectrum from the model with $v_{\infty_2}=5000\;\kmps$ relative to 
the model with $v_{\infty_2}=3000\;\kmps$. These inferences are supported
by spectra calculated with $v_{\infty_2}=5000\;\kmps$ and $\eta=0.1$, 
and variable $\Mdot_{2}$ (Fig.~\ref{fig:mdot2_var}). This shows that 
the spectrum hardens as $\Mdot_{2}$ increases from $10^{-6} \;\Msolpyr$ to
$10^{-4}\;\Msolpyr$ (the spectral index $\alpha$, where $F_{\nu} 
\propto \nu^{\alpha}$, increases by about 0.1). This is what we expect 
from our prior inferences 
since a higher $\Mdot_{2}$ helps to make postshock gas with a
preshock velocity of $v_{\infty_2}=5000 \;\kmps$ somewhat more efficient 
at radiating. In summary, we find that varying $\Mdot_{2}$ can actually have 
a small effect on the shape of the spectrum in some parts of parameter
space, as well as the more obvious and much greater effect on the 
luminosity. The power of $\Mdot_{2}$ on affecting the spectral shape is, 
however, much less than that of $v_{\infty_2}$, which is the primary 
influence.

\subsubsection{Summary of Spectral Dependence on $\eta$, $\Mdot_{2}$, and
$v_{\infty_2}$}
\label{sec:summ}

For the models investigated in this paper, we find that the spectral shape 
is i) insensitive to the value of $\eta$, ii) can harden slightly 
as $\Mdot_{2}$ increases (in other {\mbox parts} of {\mbox parameter}
space, \eg strongly radiative shocks, it may soften with increasing 
$\Mdot_{2}$), and iii) generally {\mbox hardens} with increasing 
$v_{\infty_{2}}$, 
but can sometimes soften. The {\mbox luminosity} increases with i) decreasing
$\eta$ (since $\eta$~Car is unusual in the sense that there is no
discernible contribution to the X-rays from the shocked primary wind,
bar mixing), ii) increasing $\Mdot_{2}$, and iii) increasing $v_{\infty_{2}}$.

\subsection{The Spectral Grid}
\label{sec:spec_grid}
The synthetic spectra were input into a FITS-style `Table' model
for use with the XSPEC data-analysis package, as specified by Arnaud 
(\cite{A1999}). The fitting procedure within XSPEC is able to interpolate
between the parameter values on the spectral grid so that best-fit values of
$\Mdot_{2}$ \etc are obtained from fitting the {\it Chandra} data.
In the $2-10$~keV band, the intrinsic luminosity on our grid ranges 
from $1.8 \times 10^{31} \; \ergps$ to $1.6 \times 10^{37} \; \ergps$,
which brackets the observationally determined value of 
$4 \times 10^{34} \; \ergps$ (Corcoran \etal \cite{C2001}).

\section{Fits to the {\em Chandra} grating spectrum}
\label{sec:fits}
We compared the model intrinsic colliding wind spectra with the 
observed grating spectrum allowing the 
mass-loss rate and wind terminal velocity of the companion to vary. 
We in addition included absorption from a cool overlying medium to 
simulate the combined absorption of the circumstellar (nebular) 
material in the Homunculus, the ISM, and the wind from the companion 
(all of which may contribute to the X-ray absorption at some level). 
Since the stellar wind absorption depends on the companion's mass-loss 
rate (at the phase of the {\em Chandra} observation) for consistency 
we should model this component of the absorption directly, but this 
is beyond the scope of the present paper. Because 
of the low number of counts per spectral bin in the high-resolution 
grating spectrum, and because background does not contribute significantly to 
the observed spectrum at energies above 2~keV, we fit the gross 
spectrum using the C-statistic (Cash \cite{C1979}), which is appropriate 
for Poisson-distributed data.

\begin{figure}
\begin{center}
\psfig{figure=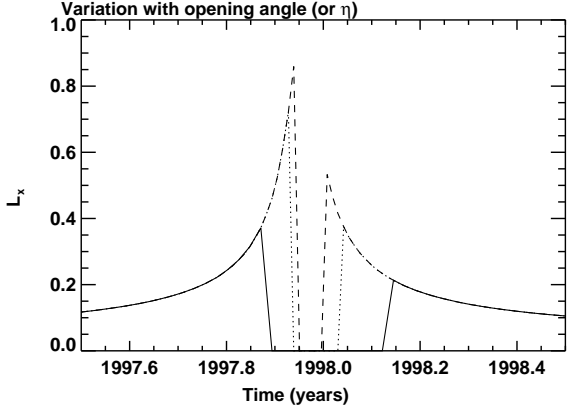,width=7.8cm}
\end{center}
\caption[]{The duration of the X-ray minimum predicted from
simple models. The wind momentum ratio in the three models shown is 
$\eta=0.2,1.0,5.0$ (solid, dots, dashes), which gives a duration of
$92, 33, 16\;{\rm d}$ respectively. The observed duration is 
$\approx 100\;{\rm d}$, so the data is best approximated by $\eta=0.2$.}
\label{fig:eta_lc2}
\end{figure}

The fact that the shapes of our model spectra are not dependent on
the value of $\eta$ introduces a {\mbox degeneracy} 
into our grid as both $\eta$ 
and $\Mdot_{2}$ primarily influence the normalization of the spectra. 
This means that various combinations of $\eta$ and $\Mdot_{2}$ can
provide similar quality {\mbox fits} to the data. However, it is possible to 
obtain an estimate for $\eta$ from the duration of the lightcurve minimum 
($\approx 100$~d). To do this we have constructed a simple 
model of the observed X-ray emission. We assume that the intrinsic flux
varies as $1/D$, and that the absorption from the shock apex along the 
line of sight is negligible when viewing through the less dense secondary
wind, but total when viewing through the very dense primary wind. We use
the latest orbital parameters (Corcoran \etal \cite{CISP2001}) and
assume that the skew angle of the shock cone from the orbital velocity,
$v_{\rm orb}$, is $\delta \simeq \arctan (v_{\rm orb}/v_{\infty_1})$.
Fig.~\ref{fig:eta_lc2} shows the results, scaled so that $L_{x}=1$ at 
periastron. The skew of the shock breaks the symmetry of the observed
emission so that the post-minimum flux is lower than the pre-minimum flux,
as observed. The duration of the minimum decreases with increasing 
$\eta$ (for $\eta = 0.1,0.2,0.3$ the duration is $133, 92, 71\;{\rm d}$),
and is best matched by 
$\eta \approx 0.2$\footnote{Incidently, the lightcurve minimum computed
with the 2D hydrodynamical model in Pittard \etal \cite{PSCI1998} 
(redisplayed in Fig.~\ref{fig:eta_lc} at the beginning of this paper) also
matches the duration of minimum very well (although not the post-minimum
flux because the skew of the shock could not be computed with a 2D
calculation). Although different mass-loss rates and a {\mbox slightly} less 
eccentric orbit were used, the wind momentum ratio adopted was also 
$\eta=0.2$, which provides a certain {\mbox robustness} for this value.}.
Therefore we adopt this value for the rest of our analysis.

\begin{figure*}
\begin{center}
\psfig{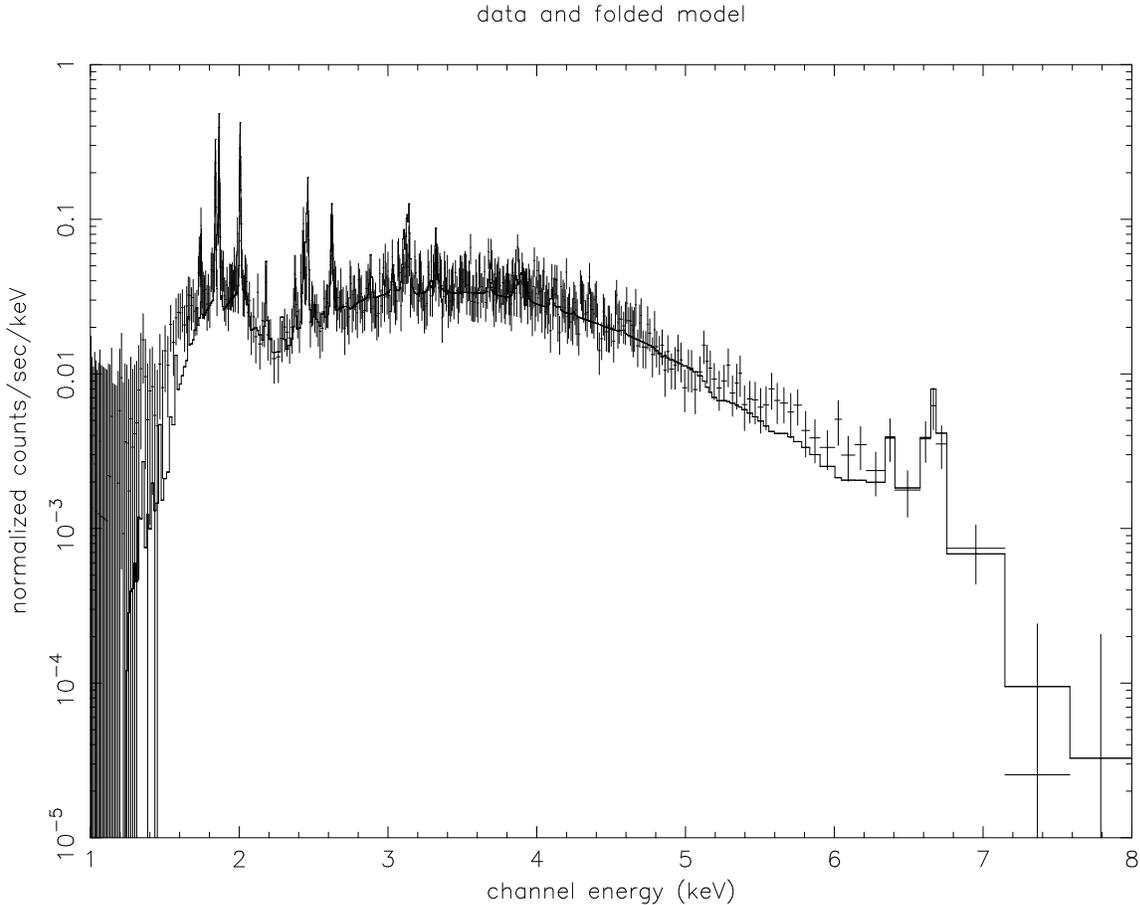}
\end{center}
\caption[]{Best fit to the {\it Chandra} grating spectrum. The fit parameters
were $\Mdot_{2} = 0.98 \times 10^{-5} \; \Msolpyr$, $v_{\infty_{2}} = 3000 
\;\kmps$, $\eta = 0.2$, $N_{\rm H} = 7.7 \times 10^{22} \; \cm^{-2}$, 
and the overall normalization was 1.16. If we take 
$v_{\infty_{1}} = 600 \;\kmps$, this implies that 
$\Mdot_{1} = 2.5 \times 10^{-4} \; \Msolpyr$. As seen from the figure, 
the continuum level is very well fitted by the interpolated model.}
\label{fig:bestfit_bb}
\end{figure*}

\begin{figure*}
\begin{center}
\psfig{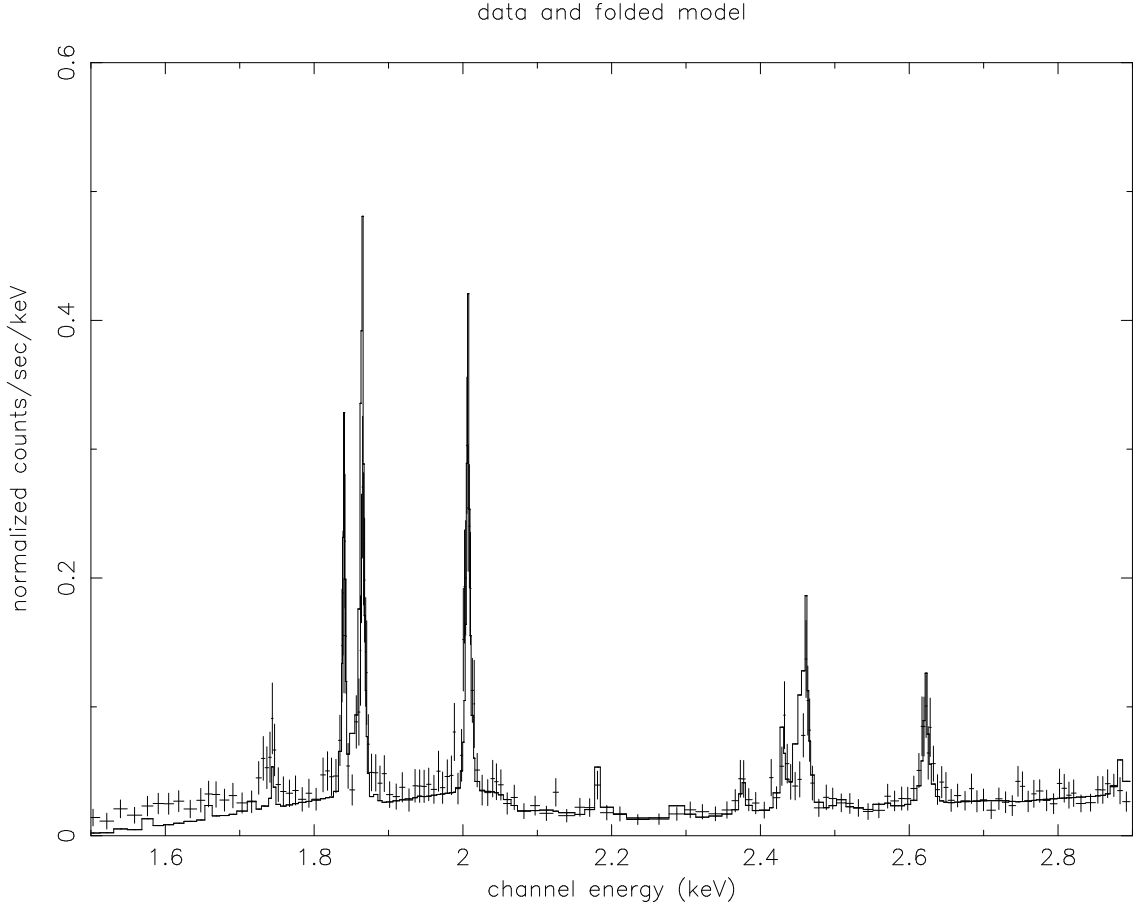}
\end{center}
\caption[]{Zoomed in view of Fig.~\ref{fig:bestfit_bb} showing the 
spectrum in the $1.8-2.8$~keV region. The lines are identified as either
S or Si, and from left to right are: Si~XIII triplet (1.839, 1.854,
1.865~keV), Si~XIV (2.006~keV), Si~XIII (2.182~keV), Si~XIV (2.375~keV),
S~XV triplet (2.429, 2.447~keV), S~XVI (2.622~keV). The fit to the S lines
is very good, while the model slightly overpredicts the emission from the Si 
lines, perhaps indicating that the Si abundance is slightly sub-solar.}
\label{fig:bestfit_lines}
\end{figure*}

We attempted to fit the grid of synthetic colliding wind spectra to the 
observed MEG~+1~order spectrum. We also constrained the spectrum 
normalization to lie between 0.8 and 1.2, which fixes the distance to 
the star between 2300 and 1900~pc, close to the canonical distance of 
2100~pc. In Fig.~\ref{fig:bestfit_bb} we show the best fit interpolated 
spectrum from our grid in the range $1.5-7.5$~keV. 
Fig.~\ref{fig:bestfit_lines} shows a close-up of the fit to the Si and S
emission lines between $1.8-2.8$~keV. Both figures demonstrate the excellent
agreement found between the data and the models and highlight the progress
made in both the observational and theoretical study of colliding winds over
the last 10 years. The continuum shape and level from the models matches
closely that of the data. All strong lines (and most weak ones) which 
appear in the observed spectrum are matched also in the best fit. This 
means that the temperature distribution seen in the grating spectrum is 
matched by the model. However most of the strong lines are significantly 
overpredicted by the model fit. This
discrepancy is in the opposite sense to the fit results from 
simple one- and two-temperature Raymond-Smith models, and is perhaps 
revealing that our assumption of solar abundances needs to be modified.
The best fit parameters are summarized in Table~\ref{tab:bestfit}.

\begin{table}
\begin{center}
\caption{The best fit parameters from the fit of the spectral grid to the
grating data. We fix the terminal velocity of the primary wind at 
$500\;\kmps$, the stellar separation at $D=4.0 \times 10^{14}\;{\rm cm}$, 
and the wind momentum ratio ($\eta$) at 0.2, and allow $\Mdot_{2}$, 
$v_{\infty_2}$, $N_{\rm H}$ and the normalization to vary. Solar abundances 
were assumed, and a cold external absorbing column was applied.
The resultant normalization is very close to 1.0, which indicates that the
model and the values of $D$ and $\eta$ are sensible. Alternatively, it
implies that our assumed distance to $\eta$~Car is too far by a factor
of 1.077, if we imagine this as the sole source of discrepancy. This would
revise the source distance down to 2.1/1.077 = 1.95~kpc, which is well
inside of observational uncertainties. The implied mass-loss 
rate of the primary star from our fit is $\Mdot_{1} = 2.5 \times 10^{-4}
\;\Msolpyr$.}
\label{tab:bestfit}
\begin{tabular}{ll}
\hline
Parameter & Value \\
\hline
$\eta$ & 0.2 (fixed) \\
$\Mdot_{2} \; (\Msolpyr)$ & $0.98^{+0.09}_{-0.08} \times 10^{-5}$  \\
$v_{\infty_{2}} \; (\kmps)$ & $3000^{+350}_{-340}$ \\
$N_{\rm H} \; ({\rm cm^{-2}})$ & $7.68^{+0.17}_{-0.17} \times 10^{22}$ \\
Normalization & 1.16 \\
\hline
\end{tabular}
\end{center}
\end{table}

\section{Conclusions}
\label{sec:conclusions}
In this paper our aim has been to test the binary hypothesis of 
$\eta$~Car by directly fitting its X-ray spectrum using a grid of spectra 
calculated from hydrodynamical models of the wind-wind collision. While our
analysis does not prove that it is a binary, we find 
that the colliding wind emission model naturally provides for the
range of ionization seen in the emission line grating spectrum for
reasonable values of the wind parameters. We have not shown that it is 
inconsistent with emission from a single star, but have noted that it is
unlike any of the other single stars observed so far at high energies and 
dispersion.

The technique applied in this paper has only been demonstrated once before
and is the first time that it has been used with a high quality grating 
spectrum. Due to the low velocity of the primary wind 
($\approx 500 \kmps$), most of the observed X-ray emission arises 
from the shocked wind of the companion star. We find it difficult 
therefore to fit both $\Mdot_{1}$ and $v_{\infty_{1}}$ as free parameters. 
However, the duration of the observed X-ray minimum can
be used to estimate the wind momentum ratio of the stars, $\eta$. 
With $\eta$ fixed at 0.2, and $\Mdot_{2}$, and $v_{\infty_{2}}$ as free 
parameters, we are able to obtain a good fit to the data.

We find that the mass-loss rate of the companion is
$\Mdot_{2} \approx 10^{-5} \;\Msolpyr$ and the terminal velocity of
{\mbox its} wind 
is $v_{\infty_{2}} \approx 3000 \; \kmps$. These values suggest 
that the companion is probably an Of supergiant (O-stars with similar 
wind parameters - \eg HD~15570 (O4If), HD~93129A (O3If), HD~93250 (O3Vf), 
HD~151804 (O8If), and Cyg~OB2~\#9 (O5If) - are listed in Howarth \& Prinja 
\cite{HP1989}), or is possibly a WR star. The velocity of the primary 
wind has been determined to 
lie in the range $500-700 \; \kmps$ (\eg Hillier \etal \cite{HDIG2001}). 
Hence our fit implies a primary mass-loss rate of 
$\Mdot_{1} \approx 2.5 \times 10^{-4} \;\Msolpyr$. From the 
uncertainty in the value
of $\eta$, and the interpolation on our grid, we estimate the uncertainty 
on our derived value for $\Mdot_{1}$ as approximately a factor of 2.
 
Our best-fit estimate of $\Mdot_{1}$ is smaller than typically inferred 
(\cf Davidson \& Humphreys \cite{DH1997}; Hillier \etal \cite{HDIG2001})
However, we note that a lower mass-loss rate for the primary star can
reduce some of the problems noted by Hillier \etal (\cite{HDIG2001}) who
fitted a value as high as $10^{-3} \;\Msolpyr$. In particular,
the models of Hillier \etal (\cite{HDIG2001}) suffered from 
absorption components that were
too strong and electron-scattering wings which were overestimated. Both 
indicate that their chosen mass-loss rate is too high. Paradoxically
both the H$\alpha$ and H$\beta$ {\mbox emission} lines were weaker than 
observed,
indicating that their mass-loss rate is too low. However, it is well known
that the wind collision zone can be a strong source of emission lines
(\eg HD~5980, Moffat \etal \cite{M1998}), which 
would resolve this problem. Their inferred minimum column density
is also larger than the observed X-ray value, again implying an 
overestimate of $\Mdot_{1}$. To increase the 
mass-loss rate of the primary towards the value estimated by 
Hillier \etal (\cite{HDIG2001}), we require either a reduction in the 
wind velocity to $\approx 100 \;\kmps$, which in turn
is in conflict with previous estimates, or a reduction in the
wind momentum ratio to $\eta \ltsimm 0.1$, which is in conflict with the
observed duration of the X-ray minimum. Finally, it is worth noting that
our results indicate a value for $\Mdot_{1}$ which 
is closer to the value inferred for the Pistol star 
($\Mdot \approx 4 \times 10^{-4} \;\Msolpyr$; Figer \etal \cite{F1998}),
an extreme early-type star with similarities to $\eta$~Car.

Since current observations and theoretical modelling of the optical 
spectrum are unable to determine the effective temperature and the stellar 
radius of the primary without first determining $\Mdot_{1}$ (\cf Hillier
\etal \cite{HDIG2001}), our independent estimate may prove to be extremely
useful in this regard. It will be interesting to see if the results from
this paper are consistent with future X-ray observations, and whether 
estimates of $\Mdot_{1}$ from observations at X-ray and other wavelengths 
can be reconciled. Future X-ray grating observations should also help
us to fix the value of the wind momentum ratio more accurately.

As the secondary wind dominates the X-ray spectrum, and its
terminal velocity appears to be high ($\approx 3000\;\kmps$), we 
should expect to see signs of Doppler broadening and shifts in the line 
profiles. While there is little evidence for this in the current spectrum, 
other orbital phases may be more favourable in this regard. The addition
of Doppler effects has already been incorporated in modelled X-ray spectra 
(Pittard \etal in preparation), and should provide further information
on wind velocities and the structure of the wind-wind collision region.  

The over-prediction of the X-ray lines in our models perhaps indicates 
that the companion has sub-solar abundances, which favours an O-type 
over a WR classification, although we would need to perform
a more detailed analysis to confirm this possibility. As the primary has
slightly enhanced abundances of C and N compared to solar, this 
suggests that to date there has been no mass exchange between the stars.
Lamers \etal (\cite{LLPW1998}) suggested that the star which dominates 
the UV GHRS spectrum is not the star which ejected the nebula since the 
abundances in the GHRS spectrum are not as evolved as the abundances in 
the nebula (which are indicative of CNO-cycle {\mbox products}). As the UV 
bright source is probably the companion 
(Hillier \etal \cite{HDIG2001}) this indicates that it was the primary
which ejected the nebula. There are also some caveats about the analysis in 
Lamers \etal (\cite{LLPW1998}) since strong C lines (which Lamers \etal took 
to indicate normal CNO abundances in the GHRS spectrum) also appear 
in stars known to be deficient in C (see the discussion in Hillier \etal
\cite{HDIG2001}). It is also interesting to note that the UV spectrum 
brightened in 1999.1~vs.~1998.2, which is suggestive of an eclipse of the UV 
source near periastron in 1998. In conclusion, the 
high energy photons (UV and X-ray) seem to be telling us about the 
companion.

We emphasize that contrary to the vast majority of colliding wind systems, 
our X-ray analysis of $\eta$~Car primarily probes the conditions of the 
shocked wind of the companion. X-ray observations of $\eta$~Car are 
therefore unique in this regard since at other wavelengths (with the 
possible exception of the far UV) the wind of the primary dominates the
observed phenomena. While our analysis has for the first time 
provided a direct estimate of the wind parameters of the companion star, 
relating these to the stellar parameters (mass, radius, luminosity) of
the companion star requires more work. It is 
anticipated that the continued multi-wavelength 
study of $\eta$~Car through and beyond the next periastron passage will 
further reveal the hidden secrets of this most enigmatic system.

\newpage
\begin{acknowledgements}
We would like to thank Keith Arnaud for help with constructing a table
model suitable for use with XSPEC. JMP would like to thank PPARC for 
the funding of a PDRA position, and MFC acknowledges that support for
this research was obtained through a cooperative agreement with NASA/GSFC:
NCC5-356. We would also like to thank the referee Kerstin Weis whose 
suggestions improved this paper. 
This work has made use of NASA's Astrophysics Data System Abstract Service.
\end{acknowledgements}

\end{document}